\documentclass[conference]{IEEEtran}
\usepackage[pdftex]{graphicx}
\usepackage{amssymb}
\usepackage{pgfplots}
\pgfplotsset{compat=newest}
\pgfplotsset{plot coordinates/math parser=false}
\newlength\figureheight
\newlength\figurewidth

\usepackage{color,soul}

\begin{document}

\title{Intra-Slot Interference Cancellation for Collision Resolution in Irregular Repetition Slotted ALOHA}
\author{\IEEEauthorblockN{G. Interdonato$^*$, S. Pfletschinger$^\dagger$, F. V\'{a}zquez-Gallego$^\dagger$, J. Alonso-Zarate$^\dagger$, G. Araniti$^*$}
\IEEEauthorblockA{$^*$University Mediterranea of Reggio Calabria, Reggio Calabria, Italy\\
giovanni.interdonato.252@studenti.unirc.it, araniti@unirc.it\\
$^\dagger$Centre Tecnol\`{o}gic de Telecomunicacions de Catalunya (CTTC), Castelldefels, Barcelona, Spain\\
\{stephan.pfletschinger, francisco.vazquez, jesus.alonso\}@cttc.es\\}}
\maketitle

\begin{abstract}
ALOHA-type protocols became a popular solution for distributed and uncoordinated multiple random access in wireless networks. However, such distributed operation of the Medium Access Control (MAC) layer leads to sub-optimal utilization of the shared channel. One of the reasons is the occurrence of collisions when more than one packet is transmitted at the same time. These packets cannot be decoded and retransmissions are necessary. However, it has been recently shown that it is possible to apply signal processing techniques with these collided packets so that useful information can be decoded. This was recently proposed in the Irregular Repetition Slotted ALOHA (IRSA), achieving a throughput $T \simeq 0.97$ for very large MAC frame lengths as long as the number of active users is smaller than the number of slots per frame. In this paper, we extend the operation of IRSA with \emph{i)} an iterative physical layer decoding processing that exploits the capture effect and \emph{ii)} a Successive Interference Cancellation (SIC) processing at the slot-level, named intra-slot SIC, to decode more than one colliding packet per slot. We evaluate the performance of the proposed scheme, referred to as Extended IRSA (E-IRSA), in terms of throughput and channel capacity. Computer-based simulation results show that E-IRSA protocol allows to reach the maximum theoretical achievable throughput even in scenarios where the number of active users is higher than the number of slots per frame. Results also show that E-IRSA protocol significantly improves the performance even for small MAC frame lengths used in practical scenarios.
\end{abstract}


\section{Introduction}

Uncoordinated Medium Access Control (MAC) protocols, such as ALOHA or Carrier Sensing Multiple Access (CSMA), are used in today's communication networks due to their capability for managing the access to a shared communication channel in a distributed manner. A clear example is the operation of the Random Access Channel (RACH) of LTE which consists in a framed slotted ALOHA scheme where slots represent orthogonal preambles, which users use to contend for the access to the resources \cite{LAYA2014}. 

Despite the congestion problems that these protocols suffer from in highly dense networks, they are still the best solutions available for completely distributed access in wireless networks. There are many scenarios where centralized-based access is not possible due to the long propagation delays (e.g. satellite communications) or due to scalability issues when the number of contending devices is extremely high and unpredictable, e.g. Machine-to-Machine (M2M) networks.

Therefore, when it comes to highly dense dynamic networks, random-based distributed protocols are the only viable solution known to date. It has been proven in the literature that, among the existing alternatives, frame-based ALOHA-type protocols can perform best when optimally configured. However, the high probability of collision will still yield low performance. To overcome this limitation, the use of Successive Interference Cancellation (SIC) techniques is becoming a hot topic in the area of MAC design.  The combination of the MAC layer with SIC techniques, traditionally employed at the PHY layer for coding purposes, is deemed to lead a major breakthrough in the performance of MAC protocols by turning collisions into useful information.

Recently, approaches based on multiple packet transmission \cite{DA} and iterative interference cancellation (IC) \cite{CRDSA}, \cite{IRSA} have shown to yield dramatic performance improvements in terms of throughput with respect to previous existing solutions. 

The Contention Resolution Diversity Slotted ALOHA (CRDSA) protocol proposed in \cite{CRDSA} was the first ALOHA-based protocol providing the adoption of SIC techniques for resolving collisions. More specifically, each packet is transmitted in two different randomly selected slots within a MAC frame. Even though this approach apparently increases the network load, it provides time-domain diversity through the transmission of a redundant copy of each packet. The replicas of each packet possess a pointer to the slot where the other replica was sent. Whenever a packet is successfully decoded, the pointer is extracted and the interference contribution caused by the twin replica on the corresponding slot is removed. The procedure is iterated, eventually permitting the recovery of the whole set of packets transmitted within the same frame. CRDSA achieves a maximum throughput, defined as probability of successful packet transmission per slot, of $T \simeq 0.55$, while the peak throughput for Framed Slotted ALOHA is just $T \simeq 0.37$. 

The CRDSA protocol was later generalized in \cite{IRSA}, allowing users to transmit more than 2 copies of the same packet per frame. In particular, the actual number of packet replicas is drawn from a probability mass function, referred to as \emph{degree distribution} \cite{IRSA}, that is optimized to achieve the maximum supportable load on the shared medium. Since the number of transmitted replicas is different from user to user, this scheme is dubbed Irregular Repetition Slotted Aloha (IRSA). In \cite{IRSA}, the operation of IRSA is described by borrowing concepts from graph codes such as belief propagation on a packet level for resolving collisions. It provides a bipartite graph representation allowing a fast analytical characterization of the IRSA performance. The convergence analysis of the SIC process shows that IRSA provides a throughput equal to $T \simeq 0.97$ if a suitable degree distribution is selected and as long as the number of available slots is greater than the number of contending devices. Despite these promising performance figures, IRSA cannot perform optimally when the number of devices is greater than the number of slots. This behavior can represent a boundary in scenarios suffering of channel overload problems such as M2M networks where a massive number of devices limits its application in realistic scenarios. 

This is the main motivation for the work presented in this paper, where we propose an extension of IRSA, referred to as Extended IRSA (E-IRSA), which can operate excellently even when the number of devices is above the number of available slots per frame. In the proposed scheme, the receiver attempts to recover as many data packets as possible for each single slot exploiting the capture effect, which enables to decode those packets received with the strongest signal in a given slot. Whenever a packet is decoded, its interference contribution is subtracted first from the overall signal received in that slot, i.e., \emph{intra-slot SIC}, and then, as well as in IRSA, from signals received in the slots where the related packet replicas have been also transmitted, i.e., \emph{inter-slot SIC}.

In summary, E-IRSA extends IRSA in two ways: $(i)$ it applies iterative physical-layer decoding that exploits the capture effect in order to decode more than one data packet per slot, and $(ii)$ it applies intra-slot SIC, in order to increase the decoding probability of the next colliding packets.

This extension has been motivated by the promising results published in \cite{MUDIRSA} and \cite{Stephan}. The work in \cite{MUDIRSA} presents a theoretical study on a generalized IRSA scheme assuming that the receiver is capable of decoding multiple colliding packets jointly using \emph{multiuser detection} (MUD) techniques in systems adopting code-division multiple access (CDMA). 

In its turn, the work in \cite{Stephan} describes a practical implementation of a further generalization of IRSA, the so-called Coded Slotted Aloha (CSA) \cite{CSA}, where several options for decoding more than one packet per slot in case of collision are considered. This work relies on concepts from physical layer network coding (PNC) \cite{PNC1}, \cite{PNC2}, and MUD and it also shows how it is possible to perform intra-slot SIC removing one or more packets from the overall signal received in a slot.

The rest of the paper is organized as follows. Section II introduces the system model and notations of E-IRSA. The description of the proposed collision resolution scheme is then provided in Section III. Simulation results are provided in Section IV. Finally, Section V concludes the paper.

\section{System Model and Notation}
\begin{figure}[!t]
\centering
\includegraphics[width=2.8in]{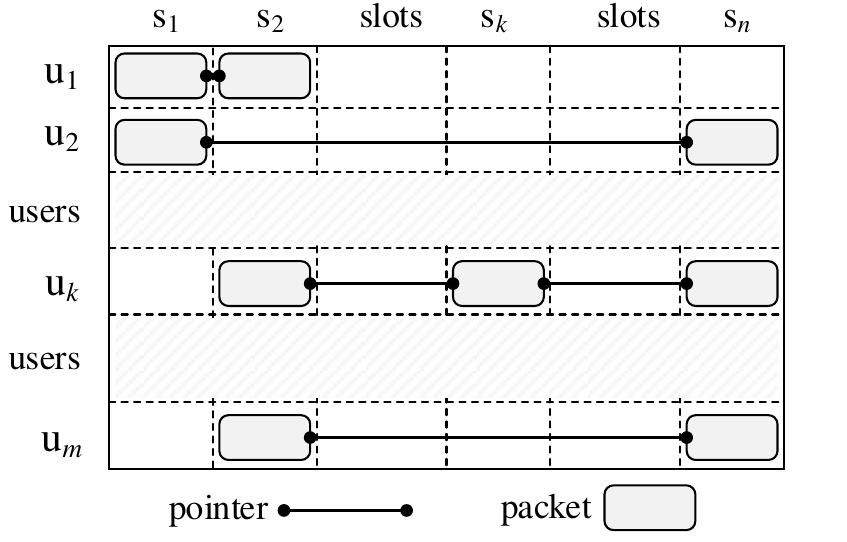}
\caption{An example of MAC frame composed by $n$ slots, with $m$ active users adopting IRSA protocol to transmit their packets.}
\label{frameSIC}
\end{figure}

\begin{figure}[!t]
\centering
\includegraphics[width=2.3in]{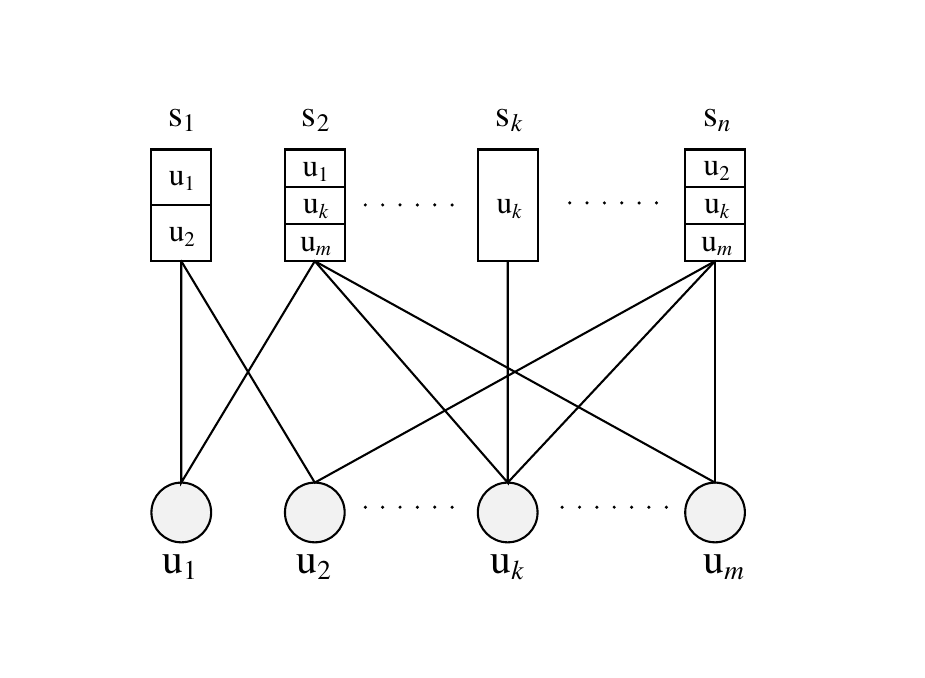}
\caption{Bipartite graph model corresponding to the above MAC frame status.}
\label{graphSIC}
\end{figure}

We consider a network composed by one receiver (also referred to as coordinator) and $m$ devices (also referred to as users) located at one-hop distance from the coordinator, forming a star topology. Every user is frame- and slot-synchronous, and has only one data packet (also referred to as burst or message) ready to transmit to the coordinator, per MAC frame. The latter is divided into $n$ slots of equal length. The transmission of a packet takes at most one slot. According to \cite{IRSA}, each of the $m$ users performs a random number of replicas of the same packet, referred to as \emph{repetition rate}, selected by a probability mass function we dubbed \emph{degree distribution}. Furthermore, the users transmit in randomly selected slots and without performing carrier sensing. Hence, each slot can be in one of three states: $(i)$ empty, i.e., no user has transmitted in the slot; $(ii)$ clean, i.e., one user has transmitted in the slot; or $(iii)$ collision, i.e., more users have transmitted in the same slot. As introduced by \cite{IRSA}, the IRSA operation can be described by a bipartite graph  $\mathcal{G}=(U,S,E)$ consisting of a set $U$ of $m$ \emph{user nodes}, i.e., one for each packet that is transmitted, a set $S$ of $n$ \emph{sum nodes}, i.e., one for each slot in the MAC frame, and a set $E$ of \emph{edges}. An edge connects a user node (UN) $u_i \in U$ to a sum node (SN) $s_j \in S$ if and only if a replica of the $i$-th packet is transmitted in the $j$-th slot. Loosely speaking, UNs correspond to packets, SNs correspond to slots, and each edge corresponds to a packet replica. Hence, a packet with $l$ replicas is represented by a UN with $l$ neighbors. A slot where $l$ replicas collide corresponds to a SN with $l$ connections. The number of edges connected to a node is referred to as the \emph{node degree}. In Fig. \ref{frameSIC} an example of IRSA MAC frame is displayed while Fig. \ref{graphSIC} shows the corresponding bipartite graph model.

\begin{figure}[!t]
\centering
\includegraphics[width=3.3in]{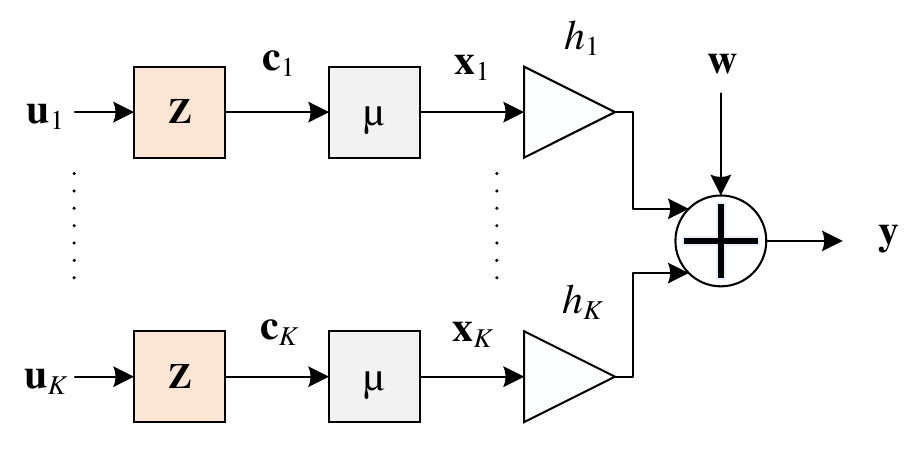}
\caption{K-user multiple-access channel with block fading. All users apply the same channel code.}
\label{channelModel}
\end{figure}

We define the \emph{logical system traffic load} $G$ as the average number of packet transmissions per slot and it is given by $G=m/n$. It provides a direct measure of the traffic handled by the scheme in contrast with the physical system load, which also takes in account the replicas. We also define the \emph{throughput} $T$ as the average number of successful packet transmission per slot and it is given by $T=G(1-P_L)$ where $P_L$ is the probability that a transmission attempt does not succeed, referred to as \emph{burst loss probability}. Hence, throughput performance depends on $G$, $P_L$ and accordingly on the variable repetition rate selected by the user. It measures the number of received packets per slot and can assume values greater than one due to the capability of receiver to decode more than one packet per slot. It is worth noticing that, this new definition of throughput is a generalization of that one provided by \cite{IRSA}, i.e., the probability of successful packet transmission per slot. The channel is modelled as block fading, where the fading coefficient $h$ follows a Rayleigh distribution with $h \sim \mathcal{CN}(0,\sqrt{\textrm{SNR}})$. It is constant during one slot, but it can vary user by user and slot by slot.
The packet of the generic user $k \in {1, 2, ..., m}$ is denoted as $\textbf{u}_k$.
Every packet is coded using a linear block code of length $C_L$ and rate $R$. The corresponding codeword symbol is $\textbf{c}_k=\textbf{u}_k\textbf{Z}$ where $\textbf{Z}$ is a common generator code matrix, which is the same for all the users. The codeword symbols are then mapped, for example, to BPSK symbols and the mapping function $\mu$ is defined element-wise as $\mu(0)=-1$, $\mu(1)=1$. Hence, the codeword symbol of a generic bit position $p$ transmitted by a generic user $k$ is given by $x_{k,p}=\mu(c_{k,p}) \in \{-1,1\}$. The choice of modulation does not change the principles of the following analysis. Assuming that in a single slot $K$ users collide, with $K \leq m$, the received signal can be written as:

\begin{equation}
\label{eq:receivedSignal}
\textbf{y} = \sum_{k=1}^K{h_k\textbf{x}_k}+\textbf{w} 
\end{equation}

where $\textbf{w}$ is the channel noise (AWGN) with $\textbf{w} \sim \mathcal{N}(0,1)$. Fig. \ref{channelModel} shows the $K$-user multiple-access channel as described above.

\section{Extended IRSA}
\begin{figure}[!t]
\centering
\includegraphics[width=2.5in]{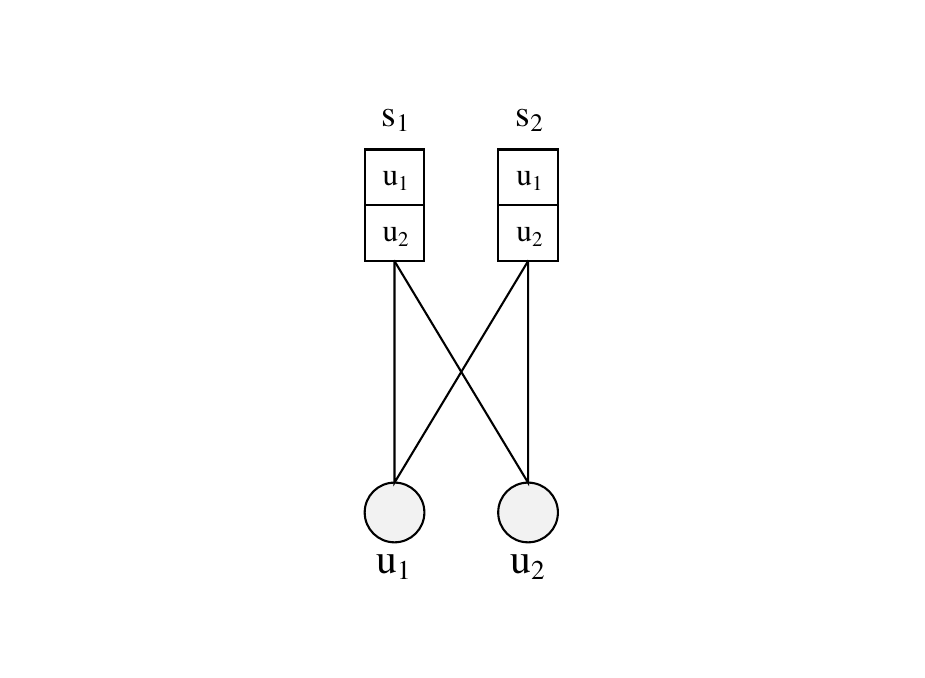}
\caption{An example of loop in the IRSA MAC frame graph representation between two user nodes and two slot nodes.}
\label{graphloop}
\end{figure}

The IRSA protocol efficiently works when a large frame size is considered and the number of active users is lower than the number of slots in the MAC frame, i.e., the logical system traffic load $G<1$. In these conditions, the probability to have a loop in the corresponding graph representation is reduced. Loop is a specific combination of collisions that interrupts the iterative IC process. For instance, there is a loop when a couple of packets are transmitted in the same couple of slots (see  Fig.\ref{graphloop}). Furthermore, the assumption on very large frame size requires a more complex receiver, introduces delay and, as a consequence, it could not be reasonable in a practical context.

Extended IRSA (E-IRSA) aims to provide an efficient collision resolution scheme by decoding multiple colliding packets per slot, by supporting logical system traffic load $G \geq 1$, and by offering optimal performance in terms of throughput, even for small MAC frame sizes. Furthermore, our work provides a complete and concrete analysis in a realistic context, such as applications for M2M scenarios.
We extend the IRSA protocol presented in \cite{IRSA} by considering the physical layer decoding combined with intra-slot and inter-slot IC processes.
More specifically, E-IRSA works as follows. Each user transmits, in randomly selected slots, a number of replica packets depending on a common probability mass function. The receiver stores the received signal of the entire frame and attempts to resolve collisions by performing a physical-layer decoding. In order to maintain low both delays and receiver complexity we considered the simplest and most ordinary decoding approach for extracting information from colliding packets. A detailed description of the adopted decoding method can be found in \cite[Section III.C]{Stephan}. 
This approach separately decodes each packet belonging to $\textbf{y}$ (i.e., the overall signal formed by colliding packets within a single slot),  considering all other packets as interference. It exploits the channel state information (CSI) of all the users and the known transmit alphabet, i.e., the BPSK constellation. The decoding process is carried out putting in a soft-input decoder, (such as a Viterbi, a turbo, or an LDPC decoder) the log-likelihood value (L-value) of the generic user $k$ and the generic bit position $p$,

\begin{equation}
\label{eq:Lvalue}
L_{k,p} \triangleq \ln \frac{P[c_{k,p}=1~|~y_p]}{P[c_{k,p}=0~|~y_p]} = \ln \frac{P[x_{k,p}=1~|~y_p]}{P[x_{k,p}=-1~|~y_p]}.
\end{equation}

For each slot by starting sequentially from the first to the last one, the receiver tries to recover as many user packets as possible, based on the received signal $\textbf{y} = [y_1, y_2, ..., y_{C_L}]$ exploiting capture effect, i.e., starting to decode from the colliding packet having the strongest signal in the slot.
Whenever a packet $\textbf{u}_{\widetilde{k}}$ is successfully decoded, the receiver can obtain the corresponding codeword $\textbf{c}_{\widetilde{k}}$ and the symbol sequence $\textbf{x}_{\widetilde{k}}$. Hence, it performs the \emph{intra-slot SIC}, subtracting the corresponding interference contribution $h_{\widetilde{k}}\textbf{x}_{\widetilde{k}}$ from the received signal $\textbf{y}$, according to

\begin{equation}
\label{eq:intraSIC}
\textbf{y} = \sum_{k=1}^K{h_k\textbf{x}_k}+\textbf{w}-h_{\widetilde{k}}\textbf{x}_{\widetilde{k}}.
\end{equation}

Then, as well as IRSA, the receiver performs the \emph{inter-slot SIC} by extracting the packet header, which contains the pointers to the slots where the replicas of the decoded packet have been transmitted, and by subtracting the interference contributions caused by the replicas, from the received signals of the corresponding slots. The IC process can be represented through a message-passing along the edges of the graph. All the details on SIC operation are described in \cite{CRDSA}, \cite{IRSA}. This procedure is iterated until the decoding process of all remaining packets of the current slot fails. When a failure occurs, E-IRSA procedure focuses to the next slot. At the end of the frame, if there are still undecoded packets, the whole E-IRSA procedure is repeated one more time.
E-IRSA procedure can be executed only under the following conditions:
\begin{itemize}
\item packets have a pointer to their $l-1$ respective replicas. The overhead due to the inclusion of pointers in the packet header may be reduced by adopting the efficient techniques described in \cite{IRSA};
\item the receiver is able to get a good estimation of the channel state information. Under this assumption the interference cancellation and the estimation of the channel parameters necessary to perform it are ideal, i.e., the receiver will estimate, as for a conventional one, the fading coefficient of every user in every slot. These assumptions simplify the analysis without substantially affecting the performance, as shown in \cite{CRDSA},\cite{IRSA}.
\end{itemize}
It is worth noticing that, differently to IRSA, E-IRSA is scalable since the presence of loops in the MAC frame graph representation does not interrupt the iterative IC process. Every loop can be resolved by performing intra-slot SIC after a successful packet decoding. In the future, we will analyze the convergence properties of the proposed IC process.

Differently to \cite{CSA}, we did not introduce an additional physical decoding step in order to minimize the user device operations with a consequence reduction of the user-side power consumption, delay and hardware complexity. 

Differently to \cite{MUDIRSA}, which considers multiuser detection techniques to jointly decode colliding packets with a fixed \emph{joint capability}, we adopt the simplest decoding approach and the number of decoded packets per slot may vary slot by slot according to the instantaneous channel quality conditions and the expected SNR value.

\section{Simulation Results}
A numerical evaluation is conducted by using MATLAB\textsuperscript{\textregistered}, to assess the effectiveness of the proposed E-IRSA. In order to make the simulative scenario as realistic as possible, our simulator includes: $(i)$ all the user-side steps from the setting of the replica repetition rate to the packet generation, $(ii)$ the generation of the pointers to the replicas, $(iii)$ a complete symbol-level implementation of the signal received by the coordinator, $(iv)$ the soft-input decoder/encoder, $(v)$ the channel model and $(vi)$ the successive interference cancellation by exploiting the pointers of the decoded packet.

\subsection{Scenario}
We implemented the channel model as block fading with additive noise. The fading coefficients, which follow a Rayleigh distribution, are constant during one slot but independent for each user and each slot. 
For all simulations, we use an LDPC code from the standard \cite{WiMAX} with code rate $R=1/2$, word length $C_L=576$ bits, and message length $RC_L=288$ bits, i.e., the packet payload size including the pointers to the other replicas. The fixed packet size for all users has been chosen just for ease of implementation. However, a variable packet size does not affect the algorithm operation. Altough the E-IRSA application is not restricted to M2M networks, it appears to be the most natural application of this scheme. For these reasons, we have taken in account a realistic packet payload size for a possible M2M scenario, e.g. smart metering.

The probability mass function, which sets the replica repetition rate of each user, is fixed to $0.5x^2+0.28x^3+0.22x^8$. According to \cite{IRSA}, this is the optimal degree distribution, for a maximum repetition rate equal to 8, that maximizes the \emph{system load threshold} $G^*$, allowing transmission with vanishing error probability ($P_L\rightarrow0$) for any offered traffic up to $G^*$, which represents the channel capacity. Finally, each simulation run has been repeated several times to get 95\% confidence intervals. 

Table \ref{tab:sim} lists the main simulation parameters related to the considered scenario. 

\begin{table}[!t]
\renewcommand{\arraystretch}{1.3}
\caption{Simulation Parameters}
\label{tab:sim}
\centering
\begin{tabular}{c c}
\hline
\bfseries Parameter & \bfseries Value\\
\hline
MAC frame length $n$ & 10, 20, 100, 1000 slots, variable\\
SNR [dB] & 10, 20, 30, 40, variable\\
Number of users $m$ & 80, 100, 120, 150, variable\\
Codeword length $C_L$ [bit] & 576\\
Code Rate $R$ & $1/2$\\
Message length $RC_L$ [bit] & 288\\
Modulation & BPSK\\
Symbol level decoder & LDPC\\
Fading & Rayleigh Block fading\\
Probability mass function & $0.5x^2+0.28x^3+0.22x^8$\\
Maximum Repetition Rate & 8 \\
\hline
\end{tabular}
\end{table}

\subsection{Performance Evaluation}
For the sake of completeness, in all the proposed numerical evaluation the performance IRSA approach are taken into account.

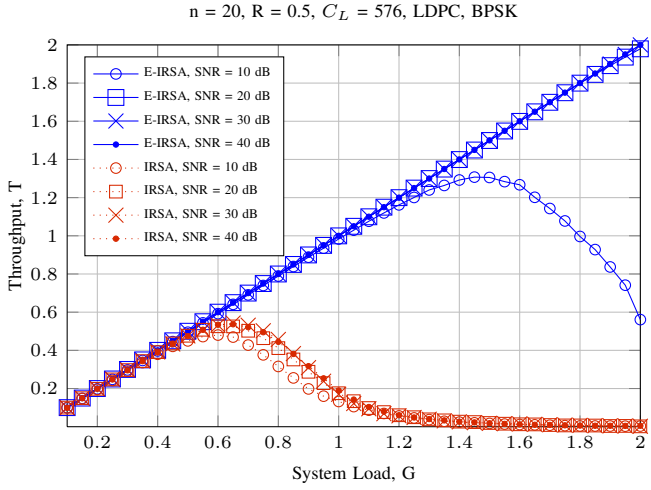
\begin{figure}[!t]
\centering
% This file was created by matlab2tikz v0.4.7 running on MATLAB 7.13.
% Copyright (c) 2008--2014, Nico Schlömer <nico.schloemer@gmail.com>
% All rights reserved.
% Minimal pgfplots version: 1.3
% 
% The latest updates can be retrieved from
%   http://www.mathworks.com/matlabcentral/fileexchange/22022-matlab2tikz
% where you can also make suggestions and rate matlab2tikz.
% 
%
% defining custom colors
\definecolor{mycolor1}{rgb}{0.84706,0.16078,0.00000}%
\begin{tikzpicture}[font=\scriptsize]

\begin{axis}[%
width=3in,
height=2in,
scale only axis,
xmin=0.1,
xmax=2,
xlabel={System Load, G},
xmajorgrids,
ymin=0,
ymax=2,
ylabel={Throughput, T},
ytick={0.2,0.4,...,2},
x label style={below=0mm},
y label style={below=-3mm},
ymajorgrids,
title={n = 20, R = 0.5, $C_L$ = 576, LDPC, BPSK},
legend style={legend pos=north west,draw=black,fill=white,legend cell align=left,font=\tiny}
]
\addplot [color=blue,solid,mark=o,mark options={solid}]
  table[row sep=crcr]{%
0.1	0.099\\
0.15	0.148333333333333\\
0.2	0.198\\
0.25	0.248\\
0.3	0.2975\\
0.35	0.3471\\
0.4	0.3968\\
0.45	0.4457\\
0.5	0.4949\\
0.55	0.5439\\
0.6	0.5922\\
0.65	0.6408\\
0.7	0.6906\\
0.75	0.7398\\
0.8	0.7893\\
0.85	0.8391\\
0.9	0.888\\
0.95	0.9351\\
1	0.9832\\
1.05	1.0294\\
1.1	1.0763\\
1.15	1.1198\\
1.2	1.162\\
1.25	1.2009\\
1.3	1.2401\\
1.35	1.2638\\
1.4	1.2923\\
1.45	1.3074\\
1.5	1.3056\\
1.55	1.2838\\
1.6	1.2673\\
1.65	1.2017\\
1.7	1.143\\
1.75	1.0778\\
1.8	0.9965\\
1.85	0.9278\\
1.9	0.8368\\
1.95	0.741333333333333\\
2	0.561\\
};
\addlegendentry{E-IRSA, SNR = 10 dB};

\addplot [color=blue,solid,mark=square,mark options={solid, scale=1.5}]
  table[row sep=crcr]{%
0.1	0.1\\
0.15	0.15\\
0.2	0.2\\
0.25	0.25\\
0.3	0.3\\
0.35	0.35\\
0.4	0.4\\
0.45	0.45\\
0.5	0.5\\
0.55	0.55\\
0.6	0.6\\
0.65	0.65\\
0.7	0.7\\
0.75	0.75\\
0.8	0.8\\
0.85	0.85\\
0.9	0.8998\\
0.95	0.9498\\
1	0.9997\\
1.05	1.0497\\
1.1	1.0997\\
1.15	1.1499\\
1.2	1.1999\\
1.25	1.2499\\
1.3	1.2999\\
1.35	1.3495\\
1.4	1.3994\\
1.45	1.4493\\
1.5	1.4992\\
1.55	1.549\\
1.6	1.5993\\
1.65	1.6494\\
1.7	1.6993\\
1.75	1.7495\\
1.8	1.7997\\
1.85	1.8455\\
1.9	1.8914\\
1.95	1.936\\
2	1.9795\\
};
\addlegendentry{E-IRSA, SNR = 20 dB};

\addplot [color=blue,solid,mark=x,mark options={solid, scale=2}]
  table[row sep=crcr]{%
0.1	0.1\\
0.15	0.15\\
0.2	0.2\\
0.25	0.25\\
0.3	0.3\\
0.35	0.35\\
0.4	0.4\\
0.45	0.45\\
0.5	0.5\\
0.55	0.55\\
0.6	0.6\\
0.65	0.65\\
0.7	0.7\\
0.75	0.75\\
0.8	0.8\\
0.85	0.85\\
0.9	0.9\\
0.95	0.95\\
1	1\\
1.05	1.05\\
1.1	1.1\\
1.15	1.15\\
1.2	1.2\\
1.25	1.25\\
1.3	1.3\\
1.35	1.35\\
1.4	1.4\\
1.45	1.45\\
1.5	1.5\\
1.55	1.55\\
1.6	1.6\\
1.65	1.65\\
1.7	1.7\\
1.75	1.75\\
1.8	1.8\\
1.85	1.85\\
1.9	1.9\\
1.95	1.95\\
2	2\\
};
\addlegendentry{E-IRSA, SNR = 30 dB};

\addplot [color=blue,solid,mark=*,mark options={solid, scale=0.5}]
  table[row sep=crcr]{%
0.1	0.1\\
0.15	0.15\\
0.2	0.2\\
0.25	0.25\\
0.3	0.3\\
0.35	0.35\\
0.4	0.4\\
0.45	0.45\\
0.5	0.5\\
0.55	0.55\\
0.6	0.6\\
0.65	0.65\\
0.7	0.7\\
0.75	0.75\\
0.8	0.8\\
0.85	0.85\\
0.9	0.9\\
0.95	0.95\\
1	1\\
1.05	1.05\\
1.1	1.1\\
1.15	1.15\\
1.2	1.2\\
1.25	1.25\\
1.3	1.3\\
1.35	1.35\\
1.4	1.4\\
1.45	1.45\\
1.5	1.5\\
1.55	1.55\\
1.6	1.6\\
1.65	1.65\\
1.7	1.7\\
1.75	1.75\\
1.8	1.8\\
1.85	1.85\\
1.9	1.9\\
1.95	1.95\\
2	2\\
};
\addlegendentry{E-IRSA, SNR = 40 dB};

\addplot [color=mycolor1,dotted,mark=o,mark options={solid}]
  table[row sep=crcr]{%
0.1	0.1\\
0.15	0.147333333333333\\
0.2	0.1955\\
0.25	0.2423\\
0.3	0.2912\\
0.35	0.336\\
0.4	0.3795\\
0.45	0.4209\\
0.5	0.4496\\
0.55	0.4721\\
0.6	0.4801\\
0.65	0.4696\\
0.7	0.4276\\
0.75	0.3764\\
0.8	0.3157\\
0.85	0.2556\\
0.9	0.198\\
0.95	0.16\\
1	0.1322\\
1.05	0.1054\\
1.1	0.0866\\
1.15	0.0693\\
1.2	0.0565\\
1.25	0.046\\
1.3	0.0373\\
1.35	0.0285\\
1.4	0.0241\\
1.45	0.0203\\
1.5	0.0172\\
1.55	0.0139\\
1.6	0.0115\\
1.65	0.0099\\
1.7	0.0079\\
1.75	0.0057\\
1.8	0.004\\
1.85	0.0037\\
1.9	0.0035\\
1.95	0.00283333333333333\\
2	0.0035\\
};
\addlegendentry{IRSA, SNR = 10 dB};

\addplot [color=mycolor1,dotted,mark=square,mark options={solid, scale=1.2}]
  table[row sep=crcr]{%
0.1	0.1\\
0.15	0.149333333333333\\
0.2	0.1985\\
0.25	0.2479\\
0.3	0.2966\\
0.35	0.3437\\
0.4	0.3908\\
0.45	0.4363\\
0.5	0.4766\\
0.55	0.5079\\
0.6	0.5261\\
0.65	0.5254\\
0.7	0.5027\\
0.75	0.4638\\
0.8	0.4117\\
0.85	0.3523\\
0.9	0.2878\\
0.95	0.2283\\
1	0.1754\\
1.05	0.1337\\
1.1	0.0978\\
1.15	0.0766\\
1.2	0.0601\\
1.25	0.0497\\
1.3	0.0386\\
1.35	0.0333\\
1.4	0.0265\\
1.45	0.0243\\
1.5	0.0195\\
1.55	0.0157\\
1.6	0.0128\\
1.65	0.0121\\
1.7	0.0089\\
1.75	0.0068\\
1.8	0.0067\\
1.85	0.0062\\
1.9	0.0038\\
1.95	0.00416666666666667\\
2	0.002\\
};
\addlegendentry{IRSA, SNR = 20 dB};

\addplot [color=mycolor1,dotted,mark=x,mark options={solid, scale=2}]
  table[row sep=crcr]{%
0.1	0.1\\
0.15	0.149333333333333\\
0.2	0.1993\\
0.25	0.2491\\
0.3	0.297\\
0.35	0.3448\\
0.4	0.3917\\
0.45	0.4333\\
0.5	0.4756\\
0.55	0.5055\\
0.6	0.5318\\
0.65	0.5431\\
0.7	0.5316\\
0.75	0.5025\\
0.8	0.457\\
0.85	0.3829\\
0.9	0.3094\\
0.95	0.2383\\
1	0.1699\\
1.05	0.1215\\
1.1	0.0939\\
1.15	0.0682\\
1.2	0.0569\\
1.25	0.0483\\
1.3	0.0392\\
1.35	0.0307\\
1.4	0.0253\\
1.45	0.0211\\
1.5	0.0177\\
1.55	0.0159\\
1.6	0.0145\\
1.65	0.0125\\
1.7	0.0102\\
1.75	0.0088\\
1.8	0.0069\\
1.85	0.0055\\
1.9	0.0047\\
1.95	0.00383333333333333\\
2	0.0025\\
};
\addlegendentry{IRSA, SNR = 30 dB};

\addplot [color=mycolor1,dotted,mark=*,mark options={solid, scale=0.5}]
  table[row sep=crcr]{%
0.1	0.099\\
0.15	0.149333333333333\\
0.2	0.199\\
0.25	0.2476\\
0.3	0.2955\\
0.35	0.3432\\
0.4	0.3886\\
0.45	0.4331\\
0.5	0.4741\\
0.55	0.5077\\
0.6	0.5343\\
0.65	0.5366\\
0.7	0.5214\\
0.75	0.495\\
0.8	0.4446\\
0.85	0.3811\\
0.9	0.3173\\
0.95	0.2535\\
1	0.1897\\
1.05	0.1422\\
1.1	0.1032\\
1.15	0.0792\\
1.2	0.0603\\
1.25	0.0483\\
1.3	0.0401\\
1.35	0.0332\\
1.4	0.0263\\
1.45	0.0227\\
1.5	0.0182\\
1.55	0.0146\\
1.6	0.012\\
1.65	0.0101\\
1.7	0.0092\\
1.75	0.0073\\
1.8	0.0059\\
1.85	0.005\\
1.9	0.0049\\
1.95	0.00433333333333333\\
2	0.005\\
};
\addlegendentry{IRSA, SNR = 40 dB};

\end{axis}
\end{tikzpicture}%
\caption{The relation throughput vs system load for IRSA and E-IRSA with MAC frame size $n=20$ and degree distribution $0.5x^2 + 0.28x^3 + 0.22x^8$. Various SNR values.}
\label{n20_TvsG}
\end{figure}

The first set of simulations we assume a fixed MAC frame size $n=20$, and a variable number of users $m$ transmitting within the MAC frame in order to have values of system traffic load $G \in [0,2]$. In Fig. \ref{n20_TvsG}, the throughput is shown by varying the system traffic load and for different SNR values: 10, 20, 30 and 40 dB, respectively.

The relation throughput versus system load provided by E-IRSA, for higher SNRs is linear, i.e., every packet is decoded, hence, the whole traffic turns into throughput. In these cases, the throughput is the maximum theoretical achievable throughput, according to the expression $T=G(1-P_L)$.

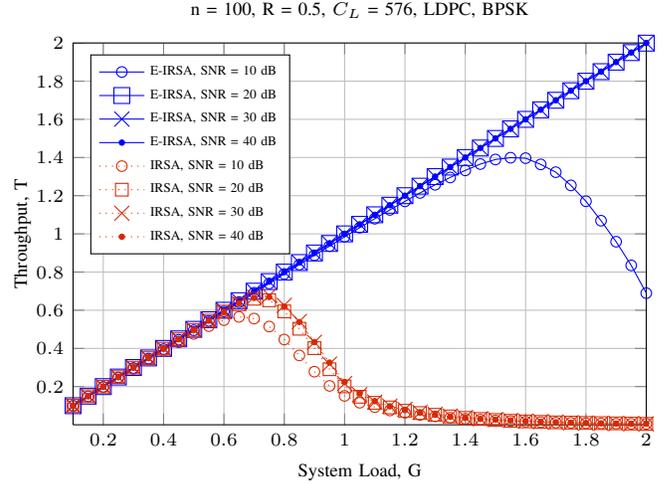
\begin{figure}[!t]
\centering
% This file was created by matlab2tikz v0.4.7 running on MATLAB 7.13.
% Copyright (c) 2008--2014, Nico Schlömer <nico.schloemer@gmail.com>
% All rights reserved.
% Minimal pgfplots version: 1.3
% 
% The latest updates can be retrieved from
%   http://www.mathworks.com/matlabcentral/fileexchange/22022-matlab2tikz
% where you can also make suggestions and rate matlab2tikz.
% 
%
% defining custom colors
\definecolor{mycolor1}{rgb}{0.84706,0.16078,0.00000}%
\begin{tikzpicture}[font=\scriptsize]

\begin{axis}[%
width=3in,
height=2in,
scale only axis,
xmin=0.1,
xmax=2,
xlabel={System Load, G},
xmajorgrids,
ymin=0,
ymax=2,
ylabel={Throughput, T},
ytick={0.2,0.4,...,2},
x label style={below=0mm},
y label style={below=-3mm},
ymajorgrids,
title={n = 100, R = 0.5, $C_L$ = 576, LDPC, BPSK},
legend style={legend pos=north west,draw=black,fill=white,legend cell align=left,font=\tiny}
]
\addplot [color=blue,solid,mark=o,mark options={solid}]
  table[row sep=crcr]{%
0.1	0.0994\\
0.15	0.148766666666667\\
0.2	0.19842\\
0.25	0.24788\\
0.3	0.29734\\
0.35	0.34676\\
0.4	0.39618\\
0.45	0.44556\\
0.5	0.49512\\
0.55	0.5442\\
0.6	0.59346\\
0.65	0.64254\\
0.7	0.69172\\
0.75	0.74032\\
0.8	0.78996\\
0.85	0.83898\\
0.9	0.88786\\
0.95	0.93598\\
1	0.98452\\
1.05	1.03224\\
1.1	1.07938\\
1.15	1.12564\\
1.2	1.17014\\
1.25	1.21434\\
1.3	1.2572\\
1.35	1.29538\\
1.4	1.33248\\
1.45	1.36564\\
1.5	1.38954\\
1.55	1.3989\\
1.6	1.39646\\
1.65	1.36386\\
1.7	1.32314\\
1.75	1.254\\
1.8	1.17074\\
1.85	1.06898\\
1.9	0.95904\\
1.95	0.8349\\
2	0.6894\\
};
\addlegendentry{E-IRSA, SNR = 10 dB};

\addplot [color=blue,solid,mark=square,mark options={solid, scale=1.5}]
  table[row sep=crcr]{%
0.1	0.0999\\
0.15	0.149966666666667\\
0.2	0.19998\\
0.25	0.25\\
0.3	0.3\\
0.35	0.35\\
0.4	0.4\\
0.45	0.45\\
0.5	0.5\\
0.55	0.55\\
0.6	0.6\\
0.65	0.64996\\
0.7	0.69988\\
0.75	0.74988\\
0.8	0.79988\\
0.85	0.84984\\
0.9	0.89982\\
0.95	0.94988\\
1	0.99988\\
1.05	1.04986\\
1.1	1.0999\\
1.15	1.14994\\
1.2	1.19996\\
1.25	1.24994\\
1.3	1.29992\\
1.35	1.3499\\
1.4	1.39984\\
1.45	1.4498\\
1.5	1.4998\\
1.55	1.5498\\
1.6	1.59978\\
1.65	1.64984\\
1.7	1.6998\\
1.75	1.74972\\
1.8	1.79974\\
1.85	1.84968\\
1.9	1.8995\\
1.95	1.94946666666667\\
2	1.999\\
};
\addlegendentry{E-IRSA, SNR = 20 dB};

\addplot [color=blue,solid,mark=x,mark options={solid, scale=2}]
  table[row sep=crcr]{%
0.1	0.1\\
0.15	0.15\\
0.2	0.2\\
0.25	0.25\\
0.3	0.3\\
0.35	0.35\\
0.4	0.4\\
0.45	0.45\\
0.5	0.5\\
0.55	0.55\\
0.6	0.6\\
0.65	0.65\\
0.7	0.7\\
0.75	0.75\\
0.8	0.8\\
0.85	0.85\\
0.9	0.9\\
0.95	0.95\\
1	1\\
1.05	1.05\\
1.1	1.1\\
1.15	1.15\\
1.2	1.2\\
1.25	1.25\\
1.3	1.3\\
1.35	1.35\\
1.4	1.4\\
1.45	1.45\\
1.5	1.5\\
1.55	1.55\\
1.6	1.6\\
1.65	1.65\\
1.7	1.7\\
1.75	1.75\\
1.8	1.8\\
1.85	1.85\\
1.9	1.9\\
1.95	1.95\\
2	2\\
};
\addlegendentry{E-IRSA, SNR = 30 dB};

\addplot [color=blue,solid,mark=*,mark options={solid, scale=0.5}]
  table[row sep=crcr]{%
0.1	0.1\\
0.15	0.15\\
0.2	0.2\\
0.25	0.25\\
0.3	0.3\\
0.35	0.35\\
0.4	0.4\\
0.45	0.45\\
0.5	0.5\\
0.55	0.55\\
0.6	0.6\\
0.65	0.65\\
0.7	0.7\\
0.75	0.75\\
0.8	0.8\\
0.85	0.85\\
0.9	0.9\\
0.95	0.95\\
1	1\\
1.05	1.05\\
1.1	1.1\\
1.15	1.15\\
1.2	1.2\\
1.25	1.25\\
1.3	1.3\\
1.35	1.35\\
1.4	1.4\\
1.45	1.45\\
1.5	1.5\\
1.55	1.55\\
1.6	1.6\\
1.65	1.65\\
1.7	1.7\\
1.75	1.75\\
1.8	1.8\\
1.85	1.85\\
1.9	1.9\\
1.95	1.95\\
2	2\\
};
\addlegendentry{E-IRSA, SNR = 40 dB};

\addplot [color=mycolor1,dotted,mark=o,mark options={solid}]
  table[row sep=crcr]{%
0.1	0.0992\\
0.15	0.148033333333333\\
0.2	0.19708\\
0.25	0.24514\\
0.3	0.29272\\
0.35	0.34062\\
0.4	0.38676\\
0.45	0.43134\\
0.5	0.47588\\
0.55	0.5167\\
0.6	0.54824\\
0.65	0.56504\\
0.7	0.55634\\
0.75	0.51462\\
0.8	0.4465\\
0.85	0.36314\\
0.9	0.27808\\
0.95	0.20374\\
1	0.1514\\
1.05	0.11556\\
1.1	0.0922\\
1.15	0.07498\\
1.2	0.06268\\
1.25	0.05166\\
1.3	0.04358\\
1.35	0.0369\\
1.4	0.0315\\
1.45	0.02714\\
1.5	0.0227\\
1.55	0.0195\\
1.6	0.01648\\
1.65	0.01374\\
1.7	0.01164\\
1.75	0.01004\\
1.8	0.00836\\
1.85	0.00692\\
1.9	0.00606\\
1.95	0.00496666666666667\\
2	0.0047\\
};
\addlegendentry{IRSA, SNR = 10 dB};

\addplot [color=mycolor1,dotted,mark=square,mark options={solid, scale=1.2}]
  table[row sep=crcr]{%
0.1	0.1\\
0.15	0.1498\\
0.2	0.19972\\
0.25	0.24964\\
0.3	0.29924\\
0.35	0.34896\\
0.4	0.3985\\
0.45	0.44798\\
0.5	0.49682\\
0.55	0.54558\\
0.6	0.59284\\
0.65	0.6314\\
0.7	0.66016\\
0.75	0.65202\\
0.8	0.59388\\
0.85	0.503\\
0.9	0.40138\\
0.95	0.29094\\
1	0.2008\\
1.05	0.14692\\
1.1	0.1151\\
1.15	0.09092\\
1.2	0.07342\\
1.25	0.06138\\
1.3	0.05232\\
1.35	0.0438\\
1.4	0.03606\\
1.45	0.03044\\
1.5	0.02516\\
1.55	0.02088\\
1.6	0.0176\\
1.65	0.01522\\
1.7	0.0127\\
1.75	0.01126\\
1.8	0.00968\\
1.85	0.00876\\
1.9	0.0076\\
1.95	0.0064\\
2	0.0049\\
};
\addlegendentry{IRSA, SNR = 20 dB};

\addplot [color=mycolor1,dotted,mark=x,mark options={solid, scale=2}]
  table[row sep=crcr]{%
0.1	0.1\\
0.15	0.149933333333333\\
0.2	0.19972\\
0.25	0.24962\\
0.3	0.29944\\
0.35	0.34906\\
0.4	0.39832\\
0.45	0.44722\\
0.5	0.49596\\
0.55	0.54374\\
0.6	0.59222\\
0.65	0.63658\\
0.7	0.66948\\
0.75	0.677\\
0.8	0.62582\\
0.85	0.54132\\
0.9	0.4323\\
0.95	0.31794\\
1	0.21618\\
1.05	0.15762\\
1.1	0.11928\\
1.15	0.09774\\
1.2	0.08114\\
1.25	0.0648\\
1.3	0.05468\\
1.35	0.04554\\
1.4	0.03756\\
1.45	0.03122\\
1.5	0.0276\\
1.55	0.02286\\
1.6	0.01918\\
1.65	0.0163\\
1.7	0.01368\\
1.75	0.01044\\
1.8	0.00884\\
1.85	0.00768\\
1.9	0.00648\\
1.95	0.006\\
2	0.0048\\
};
\addlegendentry{IRSA, SNR = 30 dB};

\addplot [color=mycolor1,dotted,mark=*,mark options={solid, scale=0.5}]
  table[row sep=crcr]{%
0.1	0.1\\
0.15	0.149866666666667\\
0.2	0.19976\\
0.25	0.24964\\
0.3	0.29942\\
0.35	0.34886\\
0.4	0.39824\\
0.45	0.44764\\
0.5	0.4967\\
0.55	0.54516\\
0.6	0.5925\\
0.65	0.63686\\
0.7	0.66438\\
0.75	0.66926\\
0.8	0.61848\\
0.85	0.5384\\
0.9	0.43322\\
0.95	0.32566\\
1	0.22364\\
1.05	0.16454\\
1.1	0.12388\\
1.15	0.09662\\
1.2	0.07742\\
1.25	0.06284\\
1.3	0.05182\\
1.35	0.0418\\
1.4	0.03634\\
1.45	0.03058\\
1.5	0.02694\\
1.55	0.02284\\
1.6	0.0192\\
1.65	0.01582\\
1.7	0.013\\
1.75	0.01098\\
1.8	0.00898\\
1.85	0.00758\\
1.9	0.00654\\
1.95	0.00573333333333333\\
2	0.0065\\
};
\addlegendentry{IRSA, SNR = 40 dB};

\end{axis}
\end{tikzpicture}%
\caption{The relation throughput vs system load for IRSA and E-IRSA with MAC frame size $n=100$ and degree distribution $0.5x^2 + 0.28x^3 + 0.22x^8$. Various SNR values.}
\label{n100_TvsG}
\end{figure}

In E-IRSA, the capability of the receiver to decode more than one packet per slot allows to manage systems with traffic loads $G \geq 1$, i.e., systems where the number of users is higher than the number of slots. This result is proved by the threshold effect occurring for $G \simeq 1.5$, even for lower SNR value such as 10 dB. Indeed, the throughput increases linearly up to the threshold then it degrades due to the threshold effect related to the iterative interference cancellation process. Similar behavior is obtained increasing the MAC frame size to $n=100$, as shown in Fig. \ref{n100_TvsG}.

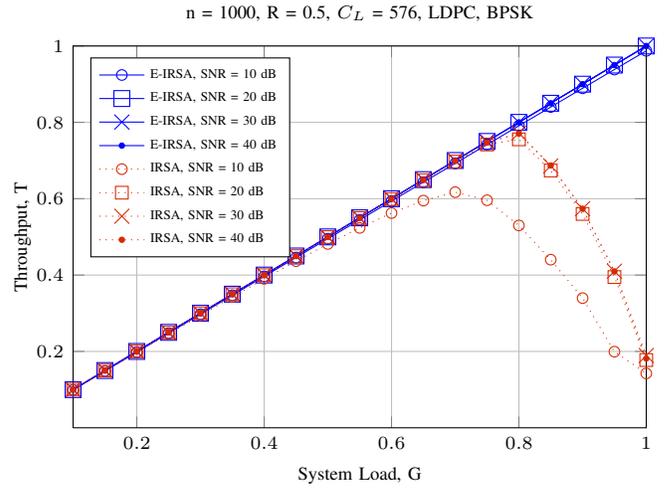
\begin{figure}[!t]
\centering
% This file was created by matlab2tikz v0.4.7 running on MATLAB 7.13.
% Copyright (c) 2008--2014, Nico Schlömer <nico.schloemer@gmail.com>
% All rights reserved.
% Minimal pgfplots version: 1.3
% 
% The latest updates can be retrieved from
%   http://www.mathworks.com/matlabcentral/fileexchange/22022-matlab2tikz
% where you can also make suggestions and rate matlab2tikz.
% 
%
% defining custom colors
\definecolor{mycolor1}{rgb}{0.84706,0.16078,0.00000}%
\begin{tikzpicture}[font=\scriptsize]

\begin{axis}[%
width=3in,
height=2in,
scale only axis,
xmin=0.1,
xmax=1,
xlabel={System Load, G},
xmajorgrids,
ymin=0,
ymax=1,
ylabel={Throughput, T},
ytick={0.2,0.4,...,1},
x label style={below=0mm},
y label style={below=-3mm},
ymajorgrids,
title={n = 1000, R = 0.5, $C_L$ = 576, LDPC, BPSK},
legend style={legend pos=north west,draw=black,fill=white,legend cell align=left,font=\tiny}
]
\addplot [color=blue,solid,mark=o,mark options={solid}]
  table[row sep=crcr]{%
0.05	0.04955\\
0.1	0.0991433333333333\\
0.15	0.148754\\
0.2	0.198304\\
0.25	0.247904\\
0.3	0.297546\\
0.35	0.347032\\
0.4	0.396482\\
0.45	0.445908\\
0.5	0.495248\\
0.55	0.544538\\
0.6	0.593896\\
0.65	0.64319\\
0.7	0.692514\\
0.75	0.74172\\
0.8	0.79097\\
0.85	0.840052\\
0.9	0.88914\\
0.95	0.93823\\
1	0.98722\\
};
\addlegendentry{E-IRSA, SNR = 10 dB};

\addplot [color=blue,solid,mark=square,mark options={solid, scale=1.5}]
  table[row sep=crcr]{%
0.05	0.05\\
0.1	0.0999933333333333\\
0.15	0.149992\\
0.2	0.19999\\
0.25	0.24998\\
0.3	0.299976\\
0.35	0.349972\\
0.4	0.399962\\
0.45	0.449954\\
0.5	0.499954\\
0.55	0.549952\\
0.6	0.599952\\
0.65	0.64995\\
0.7	0.699936\\
0.75	0.74993\\
0.8	0.79991\\
0.85	0.849906\\
0.9	0.899904\\
0.95	0.949913333333333\\
1	0.99993\\
};
\addlegendentry{E-IRSA, SNR = 20 dB};

\addplot [color=blue,solid,mark=x,mark options={solid, scale=2}]
  table[row sep=crcr]{%
0.05	0.05\\
0.1	0.1\\
0.15	0.149998\\
0.2	0.199998\\
0.25	0.249998\\
0.3	0.299998\\
0.35	0.349998\\
0.4	0.4\\
0.45	0.45\\
0.5	0.5\\
0.55	0.55\\
0.6	0.6\\
0.65	0.65\\
0.7	0.7\\
0.75	0.75\\
0.8	0.8\\
0.85	0.85\\
0.9	0.9\\
0.95	0.95\\
1	1\\
};
\addlegendentry{E-IRSA, SNR = 30 dB};

\addplot [color=blue,solid,mark=*,mark options={solid, scale=0.5}]
  table[row sep=crcr]{%
0.05	0.05\\
0.1	0.1\\
0.15	0.15\\
0.2	0.2\\
0.25	0.25\\
0.3	0.3\\
0.35	0.35\\
0.4	0.4\\
0.45	0.45\\
0.5	0.5\\
0.55	0.55\\
0.6	0.6\\
0.65	0.65\\
0.7	0.7\\
0.75	0.75\\
0.8	0.8\\
0.85	0.85\\
0.9	0.9\\
0.95	0.95\\
1	1\\
};
\addlegendentry{E-IRSA, SNR = 40 dB};

\addplot [color=mycolor1,dotted,mark=o,mark options={solid}]
  table[row sep=crcr]{%
0.05	0.04963\\
0.1	0.0988366666666667\\
0.15	0.147908\\
0.2	0.196988\\
0.25	0.245816\\
0.3	0.29434\\
0.35	0.342598\\
0.4	0.390014\\
0.45	0.436438\\
0.5	0.48149\\
0.55	0.52355\\
0.6	0.562392\\
0.65	0.594926\\
0.7	0.617182\\
0.75	0.595882\\
0.8	0.52993\\
0.85	0.440276\\
0.9	0.339612\\
0.95	0.19931\\
1	0.14238\\
};
\addlegendentry{IRSA, SNR = 10 dB};

\addplot [color=mycolor1,dotted,mark=square,mark options={solid, scale=1.2}]
  table[row sep=crcr]{%
0.05	0.05\\
0.1	0.0999933333333333\\
0.15	0.149968\\
0.2	0.19996\\
0.25	0.249914\\
0.3	0.299884\\
0.35	0.34982\\
0.4	0.399708\\
0.45	0.449578\\
0.5	0.499462\\
0.55	0.549208\\
0.6	0.599056\\
0.65	0.648448\\
0.7	0.697274\\
0.75	0.741972\\
0.8	0.755486\\
0.85	0.673876\\
0.9	0.560086\\
0.95	0.394776666666667\\
1	0.17738\\
};
\addlegendentry{IRSA, SNR = 20 dB};

\addplot [color=mycolor1,dotted,mark=x,mark options={solid, scale=2}]
  table[row sep=crcr]{%
0.05	0.05\\
0.1	0.1\\
0.15	0.149978\\
0.2	0.199956\\
0.25	0.249926\\
0.3	0.299914\\
0.35	0.3499\\
0.4	0.399868\\
0.45	0.449816\\
0.5	0.499744\\
0.55	0.549678\\
0.6	0.599516\\
0.65	0.649352\\
0.7	0.698972\\
0.75	0.747272\\
0.8	0.771916\\
0.85	0.685806\\
0.9	0.573868\\
0.95	0.410206666666667\\
1	0.18925\\
};
\addlegendentry{IRSA, SNR = 30 dB};

\addplot [color=mycolor1,dotted,mark=*,mark options={solid, scale=0.5}]
  table[row sep=crcr]{%
0.05	0.05\\
0.1	0.1\\
0.15	0.149984\\
0.2	0.199972\\
0.25	0.249964\\
0.3	0.29992\\
0.35	0.349866\\
0.4	0.399828\\
0.45	0.449754\\
0.5	0.499672\\
0.55	0.549554\\
0.6	0.599528\\
0.65	0.64925\\
0.7	0.698852\\
0.75	0.74724\\
0.8	0.77023\\
0.85	0.686974\\
0.9	0.573572\\
0.95	0.409596666666667\\
1	0.18135\\
};
\addlegendentry{IRSA, SNR = 40 dB};

\end{axis}
\end{tikzpicture}%
\caption{The relation throughput vs system load for IRSA and E-IRSA with MAC frame size $n=1000$ and degree distribution $0.5x^2 + 0.28x^3 + 0.22x^8$. Various SNR values.}
\label{n1000_TvsG}
\end{figure}

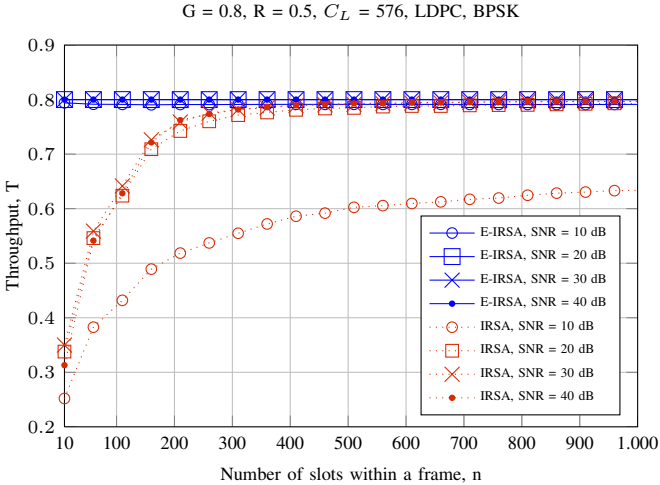
\begin{figure}[!t]
\centering
% This file was created by matlab2tikz v0.4.7 running on MATLAB 7.13.
% Copyright (c) 2008--2014, Nico Schlömer <nico.schloemer@gmail.com>
% All rights reserved.
% Minimal pgfplots version: 1.3
% 
% The latest updates can be retrieved from
%   http://www.mathworks.com/matlabcentral/fileexchange/22022-matlab2tikz
% where you can also make suggestions and rate matlab2tikz.
% 
%
% defining custom colors
\definecolor{mycolor1}{rgb}{0.84706,0.16078,0.00000}%
\begin{tikzpicture}[font=\scriptsize]

\begin{axis}[%
width=3in,
height=2in,
scale only axis,
xmin=10,
xmax=1000,
xtick={10,100,200,300,400,500,600,700,800,900,1000},
xticklabels={{10},{100},{200},{300},{400},{500},{600},{700},{800},{900},{1.000}},
xlabel={Number of slots within a frame, n},
xmajorgrids,
ymin=0.2,
ymax=0.9,
ylabel={Throughput, T},
ytick={0.1,0.2,...,0.9},
ymajorgrids,
x label style={below=0mm},
y label style={below=-3mm},
title={G = 0.8, R = 0.5, $C_L$ = 576, LDPC, BPSK},
legend style={legend pos=south east,draw=black,fill=white,legend cell align=left,font=\tiny}
]
\addplot [color=blue,solid,mark=o,mark options={solid}]
  table[row sep=crcr]{%
10	0.794\\
60	0.791439393939394\\
110	0.79114816017316\\
160	0.790378929403929\\
210	0.790988606823284\\
260	0.790830526015204\\
310	0.79102381869813\\
360	0.791140381845128\\
410	0.79124686751632\\
460	0.791183618668393\\
510	0.791240358194805\\
560	0.791132671572484\\
610	0.791088274756808\\
660	0.79118786196011\\
710	0.791094519808435\\
760	0.790962457284607\\
810	0.791021631443781\\
860	0.790902441302936\\
910	0.790915104148168\\
960	0.791052975283792\\
1010	0.791247524752475\\
};
\addlegendentry{E-IRSA, SNR = 10 dB};

\addplot [color=blue,solid,mark=square,mark options={solid, scale=1.5}]
  table[row sep=crcr]{%
10	0.8\\
60	0.8\\
110	0.799977976190476\\
160	0.799954899267399\\
210	0.79993554442869\\
260	0.799907766650912\\
310	0.799920266650912\\
360	0.799929790460436\\
410	0.799929337971748\\
460	0.799941549953315\\
510	0.799962770354043\\
560	0.799956709747983\\
610	0.799948259043757\\
660	0.799955998981838\\
710	0.799953265295771\\
760	0.799938892440262\\
810	0.799933964035333\\
860	0.799931998072892\\
910	0.799935906358458\\
960	0.799944521890376\\
1010	0.799940594059406\\
};
\addlegendentry{E-IRSA, SNR = 20 dB};

\addplot [color=blue,solid,mark=x,mark options={solid, scale=2}]
  table[row sep=crcr]{%
10	0.8\\
60	0.8\\
110	0.8\\
160	0.8\\
210	0.8\\
260	0.8\\
310	0.8\\
360	0.8\\
410	0.8\\
460	0.799996428571429\\
510	0.799996428571429\\
560	0.799996428571429\\
610	0.799996428571429\\
660	0.799996428571429\\
710	0.8\\
760	0.8\\
810	0.8\\
860	0.8\\
910	0.8\\
960	0.8\\
1010	0.8\\
};
\addlegendentry{E-IRSA, SNR = 30 dB};

\addplot [color=blue,solid,mark=*,mark options={solid, scale=0.5}]
  table[row sep=crcr]{%
10	0.8\\
60	0.8\\
110	0.8\\
160	0.8\\
210	0.8\\
260	0.8\\
310	0.8\\
360	0.8\\
410	0.8\\
460	0.8\\
510	0.8\\
560	0.8\\
610	0.8\\
660	0.8\\
710	0.8\\
760	0.8\\
810	0.8\\
860	0.8\\
910	0.8\\
960	0.8\\
1010	0.8\\
};
\addlegendentry{E-IRSA, SNR = 40 dB};

\addplot [color=mycolor1,dotted,mark=o,mark options={solid}]
  table[row sep=crcr]{%
10	0.252\\
60	0.382762626262626\\
110	0.431931980519481\\
160	0.489147365134865\\
210	0.518509730726263\\
260	0.537168821635354\\
310	0.5549197362695\\
360	0.571957831507595\\
410	0.586126760617701\\
460	0.591797728359637\\
510	0.602470679179309\\
560	0.605598173636071\\
610	0.609488314481142\\
660	0.612475053387231\\
710	0.617141720053897\\
760	0.619442025048179\\
810	0.62460056650672\\
860	0.62842917566165\\
910	0.630424459663734\\
960	0.633057561971856\\
1010	0.633831683168317\\
};
\addlegendentry{IRSA, SNR = 10 dB};

\addplot [color=mycolor1,dotted,mark=square,mark options={solid, scale=1.2}]
  table[row sep=crcr]{%
10	0.338\\
60	0.546025252525253\\
110	0.623825865800866\\
160	0.709010481185481\\
210	0.74244166398118\\
260	0.760148734688251\\
310	0.771570076151666\\
360	0.776388709692038\\
410	0.781262917836835\\
460	0.783666258850659\\
510	0.784741304388\\
560	0.78714344777307\\
610	0.7882447337498\\
660	0.788206962851967\\
710	0.789373673610345\\
760	0.789912064765511\\
810	0.790271272224718\\
860	0.791267721755235\\
910	0.792211025559299\\
960	0.793123553461115\\
1010	0.792821782178218\\
};
\addlegendentry{IRSA, SNR = 20 dB};

\addplot [color=mycolor1,dotted,mark=x,mark options={solid, scale=2}]
  table[row sep=crcr]{%
10	0.35\\
60	0.55890404040404\\
110	0.641589448051948\\
160	0.726150986513487\\
210	0.758833782212411\\
260	0.775385802414431\\
310	0.78163610729248\\
360	0.784861366091652\\
410	0.788919435473251\\
460	0.790339020726707\\
510	0.791133647320514\\
560	0.79259129698792\\
610	0.79361848621021\\
660	0.793990983630231\\
710	0.795223170579084\\
760	0.795797665432306\\
810	0.795878584513225\\
860	0.796087446015572\\
910	0.796495340752414\\
960	0.796495344932845\\
1010	0.797\\
};
\addlegendentry{IRSA, SNR = 30 dB};

\addplot [color=mycolor1,dotted,mark=*,mark options={solid, scale=0.5}]
  table[row sep=crcr]{%
10	0.313\\
60	0.541419191919192\\
110	0.628215205627706\\
160	0.721453667166167\\
210	0.762310656413479\\
260	0.77351469681752\\
310	0.781846709012641\\
360	0.786527040275581\\
410	0.789531715992022\\
460	0.791591393411377\\
510	0.793028187582597\\
560	0.793317766296566\\
610	0.794395047374337\\
660	0.794854541698382\\
710	0.795371208365049\\
760	0.795686610576258\\
810	0.796228069117716\\
860	0.796658849634148\\
910	0.796655227956191\\
960	0.796690237808122\\
1010	0.796495049504951\\
};
\addlegendentry{IRSA, SNR = 40 dB};

\end{axis}
\end{tikzpicture}%
\caption{The relation throughput vs MAC frame size for IRSA and E-IRSA with system load $G=0.8$ and degree distribution $0.5x^2 + 0.28x^3 + 0.22x^8$. Various SNR values.}
\label{G08_TvsSlot}
\end{figure}

By increasing the MAC frame length, the iterative IC process provides the best performance for $n \rightarrow \infty$. Indeed, in Fig. \ref{n1000_TvsG}, for a MAC frame size $n=1000$, IRSA scheme provides an improved throughput $T \simeq 0.8$ when $G \simeq 0.8$ and for an SNR value at least equal to 20 dB. E-IRSA scheme still provides the maximum achievable throughput as in the previous scenario. This concept is emphasized in Fig. \ref{G08_TvsSlot}.

\begin{figure}[!t]
\centering
% This file was created by matlab2tikz v0.4.7 running on MATLAB 7.13.
% Copyright (c) 2008--2014, Nico Schlömer <nico.schloemer@gmail.com>
% All rights reserved.
% Minimal pgfplots version: 1.3
% 
% The latest updates can be retrieved from
%   http://www.mathworks.com/matlabcentral/fileexchange/22022-matlab2tikz
% where you can also make suggestions and rate matlab2tikz.
% 
%
% defining custom colors
\definecolor{mycolor1}{rgb}{0.84706,0.16078,0.00000}%
\begin{tikzpicture}[font=\scriptsize]

\begin{axis}[%
width=3in,
height=2in,
scale only axis,
xmin=10,
xmax=40,
xlabel={SNR [dB]},
xmajorgrids,
ymin=0,
ymax=1.6,
ylabel={Throughput, T},
ytick={0.2,0.4,...,2},
x label style={below=0mm},
y label style={below=-3mm},
title={n = 100, R = 0.5, $C_L$ = 576, LDPC, BPSK},
legend style={legend pos=north east,draw=black,fill=white,legend cell align=left,font=\tiny}
]
\addplot [color=blue,solid,mark=square,mark options={solid}]
  table[row sep=crcr]{%
10	0.789\\
11	0.793133333333333\\
12	0.79492\\
13	0.79702\\
14	0.79822\\
15	0.79888\\
16	0.7996\\
17	0.7998\\
18	0.79988\\
19	0.79992\\
20	0.7999\\
21	0.7999\\
22	0.79994\\
23	0.79994\\
24	0.79994\\
25	0.79998\\
26	0.79998\\
27	0.79998\\
28	0.8\\
29	0.8\\
30	0.8\\
31	0.8\\
32	0.8\\
33	0.8\\
34	0.8\\
35	0.8\\
36	0.8\\
37	0.8\\
38	0.8\\
39	0.8\\
40	0.8\\
};
\addlegendentry{E-IRSA, m = 80};

\addplot [color=blue,solid,mark=x,mark options={solid}]
  table[row sep=crcr]{%
10	0.9848\\
11	0.990433333333333\\
12	0.99318\\
13	0.99598\\
14	0.99736\\
15	0.99846\\
16	0.99902\\
17	0.9995\\
18	0.99968\\
19	0.9998\\
20	0.99984\\
21	0.99982\\
22	0.99986\\
23	0.99992\\
24	0.99998\\
25	0.99998\\
26	1\\
27	0.99998\\
28	0.99998\\
29	0.99998\\
30	0.99998\\
31	0.99998\\
32	1\\
33	1\\
34	1\\
35	1\\
36	1\\
37	1\\
38	1\\
39	1\\
40	1\\
};
\addlegendentry{E-IRSA, m = 100};

\addplot [color=blue,solid,mark=*,mark options={solid}]
  table[row sep=crcr]{%
10	1.1758\\
11	1.184\\
12	1.18894\\
13	1.19348\\
14	1.19608\\
15	1.1978\\
16	1.19872\\
17	1.19922\\
18	1.19948\\
19	1.19976\\
20	1.19992\\
21	1.19996\\
22	1.19996\\
23	1.2\\
24	1.2\\
25	1.2\\
26	1.2\\
27	1.2\\
28	1.2\\
29	1.2\\
30	1.2\\
31	1.2\\
32	1.2\\
33	1.2\\
34	1.2\\
35	1.2\\
36	1.2\\
37	1.2\\
38	1.2\\
39	1.2\\
40	1.2\\
};
\addlegendentry{E-IRSA, m = 120};

\addplot [color=blue,solid,mark=+,mark options={solid}]
  table[row sep=crcr]{%
10	1.3988\\
11	1.4405\\
12	1.46046\\
13	1.48006\\
14	1.49064\\
15	1.49518\\
16	1.49768\\
17	1.49882\\
18	1.4994\\
19	1.49958\\
20	1.4997\\
21	1.49982\\
22	1.4999\\
23	1.49996\\
24	1.5\\
25	1.5\\
26	1.5\\
27	1.5\\
28	1.5\\
29	1.5\\
30	1.5\\
31	1.5\\
32	1.5\\
33	1.5\\
34	1.5\\
35	1.5\\
36	1.5\\
37	1.5\\
38	1.5\\
39	1.5\\
40	1.5\\
};
\addlegendentry{E-IRSA, m = 150};

\addplot [color=mycolor1,dotted,mark=square,mark options={solid}]
  table[row sep=crcr]{%
10	0.482966666666667\\
11	0.5116\\
12	0.550246666666667\\
13	0.57378\\
14	0.599646666666667\\
15	0.62436\\
16	0.638\\
17	0.647486666666667\\
18	0.6642\\
19	0.674966666666667\\
20	0.680746666666667\\
21	0.682366666666667\\
22	0.69048\\
23	0.694666666666667\\
24	0.696173333333333\\
25	0.69858\\
26	0.707166666666667\\
27	0.70686\\
28	0.707493333333333\\
29	0.71034\\
30	0.709526666666667\\
31	0.706213333333333\\
32	0.704826666666667\\
33	0.7037\\
34	0.70328\\
35	0.7035\\
36	0.703646666666667\\
37	0.704873333333333\\
38	0.70354\\
39	0.7005\\
40	0.696\\
};
\addlegendentry{IRSA, m = 80};

\addplot [color=mycolor1,dotted,mark=x,mark options={solid}]
  table[row sep=crcr]{%
10	0.144152777777778\\
11	0.14712037037037\\
12	0.152252777777778\\
13	0.157516666666667\\
14	0.162847222222222\\
15	0.168597222222222\\
16	0.172144444444444\\
17	0.175355555555556\\
18	0.174841666666667\\
19	0.175980555555556\\
20	0.175452777777778\\
21	0.175605555555556\\
22	0.176886111111111\\
23	0.181308333333333\\
24	0.182647222222222\\
25	0.1847\\
26	0.185738888888889\\
27	0.185402777777778\\
28	0.186541463414634\\
29	0.187441463414634\\
30	0.186741463414634\\
31	0.187174796747968\\
32	0.190844241192412\\
33	0.188505555555556\\
34	0.188147222222222\\
35	0.188527777777778\\
36	0.189897222222222\\
37	0.186441666666667\\
38	0.188275\\
39	0.189212962962963\\
40	0.193180555555556\\
};
\addlegendentry{IRSA, m = 100};

\addplot [color=mycolor1,dotted,mark=*,mark options={solid}]
  table[row sep=crcr]{%
10	0.0598\\
11	0.0625866666666667\\
12	0.064472\\
13	0.06604\\
14	0.06766\\
15	0.068368\\
16	0.06876\\
17	0.070084\\
18	0.070788\\
19	0.07112\\
20	0.071476\\
21	0.07196\\
22	0.071736\\
23	0.071484\\
24	0.0717\\
25	0.07188\\
26	0.07184\\
27	0.071744\\
28	0.072404\\
29	0.072848\\
30	0.073184\\
31	0.0734\\
32	0.07366\\
33	0.073996\\
34	0.073488\\
35	0.07354\\
36	0.073196\\
37	0.073756\\
38	0.073344\\
39	0.07336\\
40	0.07282\\
};
\addlegendentry{IRSA, m = 120};

\addplot [color=mycolor1,dotted,mark=+,mark options={solid}]
  table[row sep=crcr]{%
10	0.02116\\
11	0.0225366666666667\\
12	0.022964\\
13	0.023708\\
14	0.02412\\
15	0.02433\\
16	0.02486\\
17	0.024946\\
18	0.024964\\
19	0.025016\\
20	0.025144\\
21	0.025148\\
22	0.025274\\
23	0.02539\\
24	0.025408\\
25	0.025592\\
26	0.025428\\
27	0.025598\\
28	0.025296\\
29	0.02562\\
30	0.02533\\
31	0.02552\\
32	0.02529\\
33	0.025728\\
34	0.025528\\
35	0.025732\\
36	0.025534\\
37	0.025394\\
38	0.025096\\
39	0.0246066666666667\\
40	0.02474\\
};
\addlegendentry{IRSA, m = 150};

\end{axis}
\end{tikzpicture}%
\caption{The relation throughput vs SNR for IRSA and E-IRSA with MAC frame size $n=100$ and degree distribution $0.5x^2 + 0.28x^3 + 0.22x^8$. Various system load values.}
\label{n100_TvsSNR}
\end{figure}
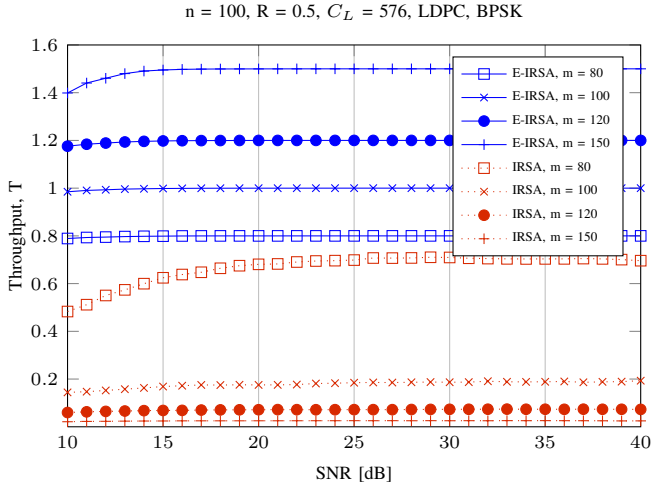

The second set of simulations assumes a fixed MAC frame size $n=100$ and evaluates the throughput by varying the number of active users $m$ (80, 100, 120, 150 users) and for different values of $SNR \in [10,40]$. The results shown in Fig. \ref{n100_TvsSNR}, highlight the opposite throughput behavior between E-IRSA and IRSA when $G$ increases. Indeed, when the number of active users per frame increases, the achievable throughput increases for E-IRSA approach, but decreases for IRSA scheme. For instance, when $G=1$, i.e., $m=100$, E-IRSA scheme provides a throughput $T=1$ and IRSA scheme provides a throughput value no greater than 0.2, while for $G=1.5$, E-IRSA and IRSA scheme provide a throughput value equal to $T \simeq 1.5$ and $T \simeq 0$, respectively. This behavior occurs when the maximum system traffic load supportable by IRSA scheme is exceeded.

The last set of simulations assumes a fixed $G$ value equal to 0.8 and evaluates the throughput by varying the MAC frame size (from 10 to 1000 slots) and the SNR values (10, 20, 30, 40 dB). Fig. \ref{G08_TvsSlot} further demonstrates that E-IRSA performance does not depend on the MAC frame size, while IRSA scheme provides the maximum throughput when $n \rightarrow \infty$.

\section{Conclusions}
The E-IRSA protocols has been presented in this paper as an extension of IRSA. By considering physical-layer decoding in combination with intra-slot and inter-slot successive interference cancellation, it is possible to decode multiple colliding packets in a given slot, thus boosting the performance of the wireless network. The results presented in this paper show that the traditional definition of throughput can reach values higher than one due to the capability of the receiver to decode more than one data packet per slot. For this reason, we have adopted the average number of successful packet transmission per slot as key performance metric, showing how E-IRSA can reach the maximum theoretical achievable throughput, i.e., the whole traffic load turns into throughput, even if the system traffic load $G \geq 1$, and independently of the MAC frame lengths and the number of contending devices. In addition, the gain in throughput leads to fewer retransmissions therefore the proposed scheme also promises improvements on device energy consumption and average delay of communications at the network level. Future work will aim at further exploring the performance of E-IRSA in terms of delay and average energy consumed by the users, separately, following the same methodology used in \cite{SIC-FSA} with respect to Framed Slotted ALOHA with SIC.

\section*{Acknowledgment}
This work has been partially funded by European Research Projects ADVANTAGE (FP7-607774) and P2P-SMARTEST (H2020-646469), and by the Catalan Government under grant (2014-SGR-1551).

\end{document}